\documentclass[11pt, epsfig]{article}
\usepackage{graphicx}
  \usepackage{amsthm,amsfonts}
  \usepackage{amsmath}
\newcommand{\bea}   {\begin{eqnarray}}
\newcommand{\eea}   {\end{eqnarray}}
\begin{document}
\renewcommand{\thefootnote}{\fnsymbol{footnote}}

\thispagestyle{empty}

\title{Pure and entangled ${\cal N}=4$ linear supermultiplets and their one-dimensional sigma-models}

\author{Marcelo Gonzales\thanks{{\em e-mail: marcbino@cbpf.br}}, Kevin Iga\thanks{{\em e-mail: Kevin.Iga@pepperdine.edu}}, Sadi Khodaee\thanks{{\em e-mail: khodaee@cbpf.br}} and Francesco
Toppan\thanks{{\em e-mail: toppan@cbpf.br}}
\\
\\
}
\maketitle

\centerline{$^{\ast}$
{\it CBPF, Rua Dr. Xavier Sigaud 150, Urca,}}{\centerline {\it\quad
cep 22290-180, Rio de Janeiro (RJ), Brazil.}
{\centerline{
$^{\dag}$
{\it Natural Science Division, Pepperdine University,}}{\centerline {\it\quad
Malibu CA 90263, USA.}}}

\maketitle
\begin{abstract}
``Pure" homogeneous linear supermultiplets (minimal and non-minimal) of the ${\cal N}=4$-Extended one-dimensional Supersymmetry Algebra are classified. ``Pure" means that they admit at least one graphical presentation (the corresponding graph/graphs are known as ``Adinkras"). \par
We further prove the existence of ``entangled" linear supermultiplets which do not admit a graphical presentation, by constructing an explicit example of an entangled ${\cal N}=4$ supermultiplet with field content $(3,8,5)$. It interpolates between two inequivalent pure ${\cal N}=4$ supermultiplets with the same field content.
The one-dimensional ${\cal N}=4$ sigma-model with a three-dimensional target based on the entangled supermultiplet is presented.\par
The distinction between the notion of equivalence for pure supermultiplets and  the notion of equivalence for their associated graphs (Adinkras) is discussed.\par
Discrete properties such as ``chirality" and ``coloring" can discriminate different supermultiplets.
The tools used in our classification include, among others, the notion of {\em field content}, {\em connectivity symbol},  {\em commuting group}, {\em node choice group} and so on.
\end{abstract}
\vfill

\rightline{CBPF-NF-008/12}

\newpage
\section{Introduction}

In this paper we present, for ${\cal N}=4$, the classification of the homogeneous (minimal and non-minimal)
linear pure supermultiplets of the
global ${\cal N}$-extended one-dimensional Supersymmetry Algebra
(the dynamical Lie superalgebra of the Supersymmetric Quantum Mechanics \cite{wit})
\bea\label{sqm}
\{Q_i, Q_j \}= 2\delta_{ij} H,&& [H, Q_i]=0,\quad i,j=1,\ldots, {\cal N}.
\eea
A pure supermultiplet admits at least one graphical presentation (the corresponding graph is known as ``Adinkra", \cite{fg}). \par
We further prove the existence of the conjectured non-adinkrizable supermultiplets \cite{dfghilm3}
(we prefer to call them here``entangled supermultiplets") which do not admit a graphical presentation,
by explicitly introducing  a supermultiplet, proved to be entangled, which interpolates between
two pure supermultiplets (the interpolation is measured by an angle). \par
We also construct a one-dimensional ${\cal N}=4$ supersymmetric $\sigma$-model with three target coordinates, defined in terms of the given entangled supermultiplet.
\par
This work is based on several advances, made in the course of the last decade, concerning the classification of
homogeneous linear supermultiplets of the one-dimensional (\ref{sqm}) ${\cal N}$-extended superalgebra \cite{pt}-\cite{top3}. The classification of the ${\cal N}=4$ pure supermultiplets is made possible by combining different tools introduced in several works \cite{kt1}-\cite{gkt}.
These tools (which include, among others, the notions of {\em engineering dimension}, {\em length of a supermultiplet}, {\em mirror symmetry duality}, {\em connectivity symbol}, {\em node choice group}, {\em commuting group}, possible {\em chirality and/or coloring of the supermultiplets}) are revised in an Appendix and further discussed in the text, when needed.\par
A key issue concerns the distinction between the equivalence class of pure supermultiplets and the equivalence class of their associated graphs (we recall that fields are represented by vertices,
while supersymmetry transformations are represented by colored edges, solid or dashed according to their sign, see the Appendix).
Equivalent graphs are related by two types of moves:
\em i}) {\em local moves}, based on the permutation of vertices with the same engineering dimension
and
\\
{\em ii}) {\em global moves}, based on the permutation of the colored edges with or without a sign flipping.\par
Global moves are responsible for properties such as the global chirality or the global color of a graph expressed by a combination of disconnected subgraphs. Indeed, if a graph is a disjoint union of disconnected subgraphs, local moves can only affect some subgraphs, while leaving unaffected the remaining ones. On the other hand, the global moves produce a global effect.
\par
Besides these two types of move, a third kind of transformation (which can be named, in reference to the famous story, a {\em gordian transformation})
can be defined. The gordian  transformation involves linear combinations of fields with the same engineering dimension. A gordian transformation produces an equivalence relation for pure supermultiplets. Under a gordian transformation a given pure supermultiplet can be associated with inequivalent (under {\em i} and {\em ii} moves) graphs. In the following we present an explicit example of a pure supermultiplet (the non-minimal ${\cal N}=4$ supermultiplet with $(4,8,4)$ field content and $8_2$ connectivity symbol) which, under a gordian transformation, can be presented either as a fully connected graph or as a disconnected graph (the disjoint union of two disconnected subgraphs).\par
The scheme of the paper is the following.
In Section {\bf 2} we review the classification of the pure supermultiplets for ${\cal N}=3$. We also discuss the notion of ``coloring" for the ${\cal N}=3$ supermultiplet with field content $(2,4,2)$. In Section {\bf 3} we present the classification of the pure (minimal and non-minimal) homogeneous linear supermultiplets for ${\cal N}=4$. They will be discriminated by their {\em field content}, {\em connectivity symbol}, {\em node choice group}, {\em commuting group}. In certain cases the notions of {\em chirality} and {\em coloring} apply.\par
In Section {\bf 4} an explicit example of an ${\cal N}=4$ supermultiplet which cannot be realized as a pure supermultiplet is constructed (``entangled supermultiplet"). It corresponds to an interpolation, with a given angle,
of two pure ${\cal N}=4$ supermultiplets with field content $(3,8,5)$.
In Section {\bf 5} we construct the one-dimensional sigma-model with a three-dimensional target which possesses a manifest ${\cal N}=4$ off-shell invariant action under the entangled supermultiplet of Section {\bf 4}.
In the Conclusions we make further comments on our results and discuss the future perspectives of our work. In order to make the paper self-contained, the needed definitions,
tools and conventions are recalled in an Appendix.

\section{The ${\cal N}=3$ supermultiplets}

We present at first the list of the ${\cal N}=3$ minimal supermultiplets. This is a necessary preliminary step for producing minimal and non-minimal ${\cal N}=4$ pure supermultiplets.
Indeed, these ones are obtained from the ${\cal N}=3$ supermultiplets by adding a compatible fourth supertransformation (in the case of a non-minimal ${\cal N}=4$ supermultiplet two separate ${\cal N}=3$ supermultiplets are employed).\par
All minimal ${\cal N}=3$ supermultiplets are pure supermultiplets associated with a graphical presentation. They contain $4$ bosonic and $4$ fermionic component fields. Their list specifying their properties
(field content $F.C.$, connectivity symbol $C.S.$, commuting group $C.G.$, node choice group $N.C.G.$ and coloring $col.$, see the Appendix) is the following
\begin{center}
\begin{tabular}{|c|c|c|c|c|}\hline
\emph{F.C.}& \emph{C.S.}&\emph{C.G.}& \emph{N.C.G.} & \emph{col.}   \\
\hline
$(4,4)$&  $4_0$& $SU(2)$ & $\{evens \}$ & $1$   \\
\hline
$(1,4,3)$ & $1_3 +3_2$& ${\bf 1}$ & $<000>$ &$1$ \\ \hline
 $(2,4,2)$&  $2_2 + 2_1$ &
$U(1)$ & $<110>$ & $3$  \\
 \hline
$(3,4,1)$&  $3_1 + 1_0$& ${\bf 1}$ & $<000>$ & $1$ \\
\hline $(1,3,3,1)$ &     &   & $<000>$   & $1$
\\\hline
\end{tabular}
\end{center}
\begin{eqnarray}&&\end{eqnarray}
(where ``$\{evens\}$" denotes the set containing the words with even number of $1$'s).\par
The above supermultiplets come out in two variants (bosonic or fermionic) according to the grading
(even or odd) of the component fields with lowest engineering dimension (in application to supersymmetric models the fermionic version of the $(4,4)$ root supermultiplet is often denoted as ``$(0,4,4)$").\par
The above supermultiplets are non-chiral (see the discussion in the Appendix). \par
The $(2,4,2)$ supermultiplet appears in $3$ different colorings (see the Appendix), related to its presentation in terms of the node choice group ($<110>$, $<101>$ or $<011>$). The three colorings are related by global moves (permutation of the supertransformations), so that, under
global moves, the three colorings belong to the same class of equivalence. On the other hand,
the modulus $C$ introduced in the Appendix ($C=|\sum_{r=1}^n j_r|$) specifies the different classes of equivalence, under local and global moves, of a collection of $n$ independent $(2,4,2)$ ${\cal N}=3$ supermultiplets, each one characterized by its coloring $j_r$.
In this construction the three colorings are put in correspondence with the three third roots of unity ($j_r^3=1$, for $r=1,2,3$).
The graph characterizing the collection of the $n$ independent supermultiplets is the disjoint union of the $n$ graphs associated to the  independent minimal pure supermultiplets.  \par
For clarity we present the graphs (Adinkras) associated to the three colorings of the $(2,4,2)$
${\cal N}=3$ supermultiplet. The three supertransformations are painted in black, red and green.
The graphs are presented in Figure {\bf 1}, {\bf 2} and {\bf 3}.

\begin{figure} [h!]
  \centering
    \includegraphics[width=0.35\textwidth]{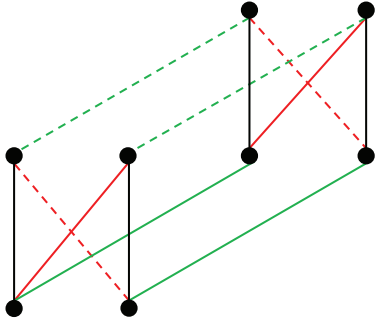}
     \caption{$N.C.G. <110>$}
    \label{figure1}
\end{figure}

\begin{figure} [h!]
  \centering
    \includegraphics[width=0.35\textwidth]{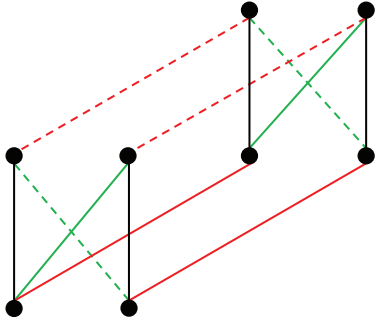}
     \caption{$N.C.G. <101>$}
    \label{figure2}
\end{figure}

\begin{figure} [h!]
  \centering
    \includegraphics[width=0.35\textwidth]{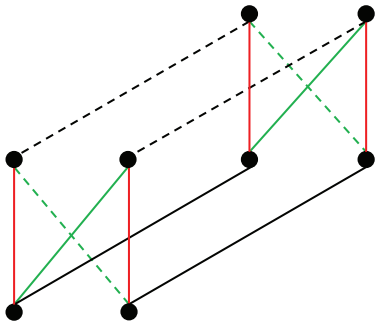}
     \caption{$N.C.G. <011>$}
    \label{figure3}
\end{figure}

\section{The classification of pure ${\cal N}=4$ supermultiplets}

The minimal ${\cal N}=4$ supermultiplets contain $4$ bosonic and $4$ fermionic fields. They are all pure supermultiplets associated with a graphical presentation. Their complete list, together with their properties, is given by the table below (one should note that, among the ${\cal N}=3$ supermultiplets, the length-$4$ $(1,3,3,1)$ is the only one which cannot be extended to ${\cal N}=4$ by adding a compatible fourth supertransformation)
\begin{center}
\begin{tabular}{|c|c|c|c|c|}\hline
\emph{F.C.}& \emph{C.S.}&\emph{C.G.}& \emph{N.C.G.} & \emph{col.}   \\
\hline
$(4,4)$&  $4_0$& $SU(2)$ & $\{evens \}$ & $1$   \\
\hline
$(1,4,3)$ & $4_3$& ${\bf 1}$ & $<0000>$ &$1$ \\ \hline
 $(2,4,2)$&  $4_2$ &
$U(1)$ & $<1100,0011> $ & $3$  \\
 \hline
$(3,4,1)$&  $4_1$& ${\bf 1}$ & $<0000>$ & $1$
\\\hline
\end{tabular}
\end{center}
\begin{eqnarray}&&\end{eqnarray}
The above supermultiplets come out in four variants. Similarly to the ${\cal N}=3$ supermultiplets
they are either bosonic or fermionic. Unlike the ${\cal N}=3$ supermultiplets they are either chiral or antichiral ($\eta=\pm 1$, see the Appendix). The chirality is flipped by global moves
so that, for a single supermultiplet, there is only one class of equivalence under global moves. For a collection of $n$ independent minimal ${\cal N}=4$ supermultiplets their overall chirality is discriminated by the modulus $\Delta=|\sum_{r=1}^n\eta_r|$ which defines the classes of equivalence under local and global moves (see the Appendix).

\subsection{Non-minimal, pure, disconnected supermultiplets}

The non-minimal, pure, ${\cal N}=4$ supermultiplets consists of $8$ bosonic and $8$ fermionic fields (that is, twice the number of the component fields of the  minimal ${\cal N}=4$ supermultiplets). Two classes of graphs are associated to the non-minimal, pure supermultiplets: the disconnected graphs, obtained by the disjoint union of $2$ graphs associated with the minimal supermultiplets, and the connected graphs (presenting a path of colored edges connecting each vertex to any other vertex). The complete list of disconnected graphs for non-minimal, pure supermultiplets of length $l=2,3$ is presented in the table below. The data reported in the table are the field content ($F.C.$), the connectivity symbol ($C.S.$), the commuting group ($C.G.$), the node choice group ($N.C.G.$), together with the decomposition into direct sum of minimal ${\cal N}=4$ supermultiplets and the label used to name the graphs (``$FR$" stands for ``Fully Reducible"). We have
\begin{center}
\begin{tabular}{|c|c|c|c|c|c|}\hline
\emph{F.C.}& \emph{label}& \emph{decomposition}&\emph{C.S.} & \emph{C.G.} & \emph{N.C.G.}\\
\hline $(8,8)$ & $FR$  & $(4,4)\oplus(4,4)$  &  $8_{0}$   &
$SU(2)\otimes SU(2)\otimes \mathbb{R}$ &  $\{evens\}$
\\\hline $(1,8,7)$&$FR$& $(1,4,3)\oplus (0,4,4)$ &$4_4 + 4_3$ & $SU(2)\otimes {\bf 1}_2 \otimes
\mathbb{R}$& $<1111>$
\\\hline
$(2, 8, 6)$& $a$&$(2,4,2)\oplus(0,4,4)$& $4_4 + 4_2$ &    $SU(2)\otimes U(1)\otimes \mathbb{R}$ & $<1100,0011>$        \\
           & $b$& $(1,4,3)\oplus(1,4,3)$&$8_3$ &   ${\bf
1}_2\otimes {\bf 1}_2 \otimes \mathbb{R}$        &     $<1111>$     \\
\hline $(3, 8, 5)$ &$a$&$(3,4,1)\oplus(0,4,4)$ &$4_4 +4_1$&
$SU(2)\otimes {\bf 1}_2\otimes \mathbb{R}$        &     $<1111>$     \\
            &$b$&$(2,4,2)\oplus(1,4,3)$& $4_3 + 4_2$&    $U(1)\otimes {\bf 1}_2 \otimes \mathbb{R}$ &  $<1111>$     \\ \hline
$(4, 8, 4)$& $a$& $(4,4,0)\oplus(0,4,4)$&$4_4 + 4_0$ &    $SU(2)\otimes SU(2)\otimes \mathbb{R}$       &  $\{evens\}$        \\
           &$b$& $(3,4,1)\oplus(1,4,3)$&$4_3+4_1$ &   ${\bf 1}_2\otimes
{\bf 1}_2 \otimes \mathbb{R}$        &   $<1111>$       \\
           &$c$&$(2,4,2)\oplus(2,4,2)$ &$8_2$ &     $U(1)\otimes U(1)\otimes \mathbb{R}$      &  $<1100,0011>$        \\
          &$d$&$(2,4,2)\oplus(2,4,2)$ &$8_2$ &     $U(1)\otimes U(1)\otimes \mathbb{R}$      &  $<1111>$        \\\hline
$(5, 8, 3)$& $a$& $(4,4,0)\oplus(1,4,3)$&$4_4+4_3$ &  $SU(2)\otimes {\bf 1}_2 \otimes \mathbb{R}$         & $<1111>$         \\
           & $b$& $(3,4,1)\oplus(2,4,2)$&$4_2 + 4_1$ &  $U(1)\otimes{\bf 1}_2 \otimes \mathbb{R}$         & $<1111>$         \\ \hline
$(6, 8, 2)$& $a$&$(4,4,0)\oplus(2,4,2)$& $4_2 + 4_0$ &   $SU(2)\otimes U(1)\otimes \mathbb{R}$        &  $<1100,0011>$       \\
           & $b$& $(3,4,1)\oplus(3,4,1)$&$8_1$  &   ${\bf 1}_2\otimes{\bf 1}_2 \otimes \mathbb{R}$        &    $<1111>$     \\ \hline
$(7, 8, 1)$ &$FR$& $(4,4,0)\oplus(3,4,1)$&$4_1 + 4_0$ &
$SU(2)\otimes {\bf 1}_2 \otimes \mathbb{R}$& $<1111>$\\ \hline
\end{tabular}
\end{center}
\begin{eqnarray}\label{n4disc}
&&
\end{eqnarray}
The above supermultiplets come out in four variants: either bosonic or fermionic and either chiral
($\Delta=2$) or non-chiral ($\Delta=0$). For $\Delta=2$ ($\Delta=0$) the supermultiplet is decomposed into $2$ minimal supermultiplets of same (opposite) chirality. The suffix $\Delta$ is used to discriminate the two inequivalent cases (therefore, the non-chiral supermultiplet $(3,8,5)_b$ will be denoted as ``$(3,8,5)_{b,\Delta=0}$").\par
The supermultiplets $(4,8,4)_c$ and $(4,8,4)_d$ possess a different node choice group. This is
a consequence of the decomposition into two minimal $(2,4,2)$ supermultiplets with same coloring (in the  $(4,8,4)_c$ case) or different coloring (in the $(4,8,4)_d$ case). \par
In terms of node choice group presentations (colorings), the supermultiplets in (\ref{n4disc}) admitting inequivalent (under local moves) presentations are
$$(2,8,6)_a,\quad (4,8,4)_c,\quad (6,8,2)_a,$$
coming out in $3$ colorings belonging to the same equivalence class under global moves.
\par
The commuting group of the supermultiplets in (\ref{n4disc}) is the tensor products of three groups: the two independent commuting groups acting on the left (right)  minimal supermultiplets of the decomposition and ${\mathbb{R}}$, acting as $+{\mathbb{Id}}$ on the component fields of the left minimal supermultiplet and as $-{\mathbb{Id}}$ on the component fields of the right minimal supermultiplet. \par
By suitably adjusting the relative engineering dimension of the two minimal supermultiplets,
non-minimal disconnected supermultiplets with length $l>3$ can be constructed. We just mention two examples: the length-$4$ supermultiplet with field content
\begin{eqnarray}
(4,4,4,4) &=& (4,4,0,0)\oplus (0,0,4,4),
\end{eqnarray}
which comes out in four variants (either bosonic or fermionic and either chiral or non-chiral)
and the interesting case of the supermultiplet with field content
\begin{eqnarray}
(2,6,6,2) &=& (2,4,2,0)\oplus (0,2,4,2),
\end{eqnarray}
which comes out in $8$ variants, namely either bosonic or fermionic, either chiral or non-chiral,
either with $C=2$ (same coloring of the $(2,4,2)$ minimal supermultiplets) or $C=1$ (different
coloring of the $(2,4,2)$ minimal supermultiplets).
\subsection{Non-minimal pure supermultiplets with a connected graph}

A graph is connected if any given vertex is connected to any other vertex by a path of colored edges
representing the supersymmetry transformations of the component fields.
The list, together with their properties, of the non-minimal ${\cal N}=4$ pure supermultiplets of length $l=2,3$
represented by a connected graph is given in the table below. All these supermultiplets are non-chiral and can be {\em oxidized}
(in the language of \cite{grt}) to ${\cal N}=8$ (i.e., four extra supersymmetry transformations can be consistently introduced so that
${\cal N}=8$ is the maximal number of supersymmetries acting on the given component fields). We have

\begin{center}
\begin{tabular}{|c|c|c|c|c|}\hline
\emph{F.C.}& \emph{label:}& \emph{C.S.}&\emph{C.G.}& \emph{N.C.G.} \\
\hline $(8,8)$ & $conn.$ & $8_{0}$ & $SU(2)\otimes
SU(2)\otimes\mathbb{R}$ & $\{evens\}$
\\\hline
$(1,8,7)$ & $conn.$ & $4_4 +4_3$ & ${\bf 1}$  &  $<0000>$ \\
\hline $(2, 8, 6)$& $A$& $2_4 + 4_3 + 2_2$& $ U(1)$ & $<1100>$  \\
\cline{2-5} & $B$& $8_3$ & $\mathbb{R}$ & $<1111>$  \\ \hline $(3,
8, 5)$ &$A$& $1_4 +3_3 + 3_2 + 1_1$& ${\bf 1}$ & $<0000>$  \\ \cline
{2-5} &$B$& $4_3 + 4_2$& ${\bf 1}$
 & $<0000>$ \\ \hline $(4, 8, 4)$& $A$& $1_4 + 6_2 + 1_0$ &
${\bf 1}$ & $<0000>$ \\ \cline {2-5} &$B$& $4_3 + 4_1$ & $SU(2)$ & $<0110,0101>$\\
\cline {2-5} &$C$& $2_3 + 4_2 +2_1$ & ${\bf 1}$ & $<0000>$ \\ \cline
{2-5} &$D$& $8_2$ & $U(1)\otimes U(1)\otimes \mathbb{R}$ & $<1100,0011>$ \\
\hline $(5,
8, 3)$& $A$& $1_3 +3_2 + 3_1 + 1_0$& ${\bf 1}$ & $<0000>$ \\
\cline{2-5} & $B$& $4_2 + 4_1$ & ${\bf 1}$
 & $<0000>$
\\ \hline $(6, 8, 2)$& $A$& $2_2 + 4_1 + 2_0$ & $U(1)$ & $<1100>$ \\
\cline{2-5} & $B$& $8_1$ & $\mathbb{R}$ & $<1111>$  \\ \hline $(7,
8, 1)$ &$conn.$& $4_1 + 4_0$& ${\bf 1}$ & $<0000>$ \\ \hline
\end{tabular}
\end{center}
\begin{eqnarray}\label{conngraph}
&&
\end{eqnarray}

The above supermultiplets are either bosonic or fermionic.\par
Inequivalent presentations of the node choice group under local moves (inequivalent colorings) are encountered
in the following cases:\par
$(2,8,6)_{A}$ and $(6,8,2)_A$ ($6$ colorings),
$(4,8,4)_{B}$ ($12$ colorings) and
$(4,8,4)_{D}$ ($3$ colorings).\par
For higher length ($l=4,5$), the list of ${\cal N}=4$ non-minimal supermultiplets with a connected graph is
given by the table
\begin{center}
\begin{tabular}{|c|c|}\hline
\emph{field content:}&${\cal N}_{max}$\emph{:}\\ \hline $(1,4,6, 4,
1)$ & $4$ \\ \hline $(1,4, 7, 4)\leftrightarrow (4, 7, 4, 1)$ & $4$
\\ \cline{1-1}\hline  $(1 ,5, 7, 3)\leftrightarrow (3, 7, 5, 1)$ & $5$ \\
\cline{1-1}\hline $(1, 6, 7, 2)\leftrightarrow (2, 7, 6, 1)$ & $5$
\\ \cline{1-1}
\hline $(2, 6, 6, 2)$ &
$6$ \\ \hline$(1, 7, 7,1)$ & $7$ \\  \hline
\end{tabular}
\end{center}
\begin{eqnarray}\label{l45}
&
&
\end{eqnarray}
which reports their field content and  the maximal number ${\cal N}_{max}$ of their oxidized
supersymmetry. The supermultiplets connected by arrows are dual under mirror symmetry (see the Appendix),
while the remaining ones are self-dual. The above supermultiplets are non-chiral and appear in two variants (bosonic or fermionic).\par
The tables (\ref{conngraph}) and (\ref{l45}) present the complete list of ${\cal N}=4$ non-minimal supermultiplets associated to a connected graph.\par
Some remarks should be made. The notion of {\em connectivity symbol} allows to discriminate inequivalent supermultiplets possessing the
same field content, commuting group and node choice group. Indeed, if we compare the $(4,8,4)_A$ with the $(4,8,4)_C$ supermultiplet we notice that
their only difference lies in their respective connectivity symbol.\par
Four cases in table (\ref{conngraph}), involving the supermultiplets $(8,8)_{conn}$, $(2,8,6)_B$, $(4,8,4)_D$ and $(6,8,2)_B$, are particularly intriguing.
For each such supermultiplet a related non-chiral pure supermultiplet entering table
(\ref{n4disc})  (that is, admitting a disconnected graph presentation) and possessing the same
 field content, connectivity symbol, commuting group and node choice group, can be found. We have the equivalence
\begin{eqnarray}\label{conndiscon}
\begin{array}{|ccc|}\hline
   Connected: &                  & Disconnected:\\\hline
  (8,8)_{conn}   & \Leftrightarrow & (8,8)_{FR,\Delta=0}  \\ \hline
  (2,8,6)_{B} & \Leftrightarrow &(2,8,6)_{b,\Delta=0}\\ \hline
  (4,8,4)_{D} & \Leftrightarrow & (4,8,4)_{c,\Delta=0}\\\hline
  (6,8,2)_{B} & \Leftrightarrow & (6,8,2)_{b,\Delta=0}\\ \hline
\end{array}
\end{eqnarray}
It is possible to prove, under general considerations, that identical properties
(field content, connectivity symbol, commuting group and node choice group) shared by inequivalent graphs imply that they represent the same supermultiplet
(as far as supersymmetry transformations are concerned). A supermultiplet can be associated with inequivalent graph presentations (the group of equivalence
for graphs is based on the local and global moves discussed in the Introduction). The gordian transformations presented in the Introduction induce an equivalence
relation for supermultiplets. They cannot be regarded, however, to be an equivalence relations for the graphs.\par
The four equivalence relations expressed in (\ref{conndiscon}) admit explicit gordian transformations which allow ``cutting" the connected graphs on the
left into the union of two separate disconnected graphs on the right.

\subsection{A gordian transformation}

We present here, explicitly, the gordian transformation involving the supermultiplets in (\ref{conndiscon}) with field content $(4,8,4)$
(the three remaining cases in (\ref{conndiscon}) admit similar gordian transformations).
Let $(\upsilon_0,\upsilon_1,{\bar \upsilon}_0,{\bar\upsilon_1};\lambda_0,\lambda_1,\lambda_2,\lambda_3,{\bar\lambda}_0,{\bar\lambda}_1,{\bar\lambda}_2,{\bar\lambda_3};g_2,g_3,{\bar g}_2,{\bar g}_3)$ be the component fields associated to the connected $(4,8,4)_D$ graph and
$(u_0,u_1,{\bar u}_0, {\bar u}_1; \psi_0,\psi_1,\psi_2,\psi_3,{\bar\psi}_0,{\bar\psi}_1,{\bar\psi}_2,{\bar\psi}_3; f_2,f_3,{\bar f}_2,{\bar f}_3)$ the component fields
associated to the disconnected non-chiral graph $(4,8,4)_{c,\Delta=0}$. The four supertransformations (painted in blue, red, green and yellow) are directly read from Figure {\bf 4}.
The fields of lowest engineering dimension are related through the gordian transformation
\begin{eqnarray}\label{gord}
&u_0=\upsilon_0-{\bar\upsilon}_0,\quad u_1=\upsilon_1-{\bar\upsilon}_1,\quad {\bar u}_0= \upsilon_0+{\bar\upsilon}_0,\quad {\bar u}_1=\upsilon_1+{\bar\upsilon}_1&
\end{eqnarray}
(the gordian transformations relating the fields with higher engineering dimension are directly read from (\ref{gord})).

\begin{figure}[h]
  \centering
    \includegraphics[width=0.85\textwidth]{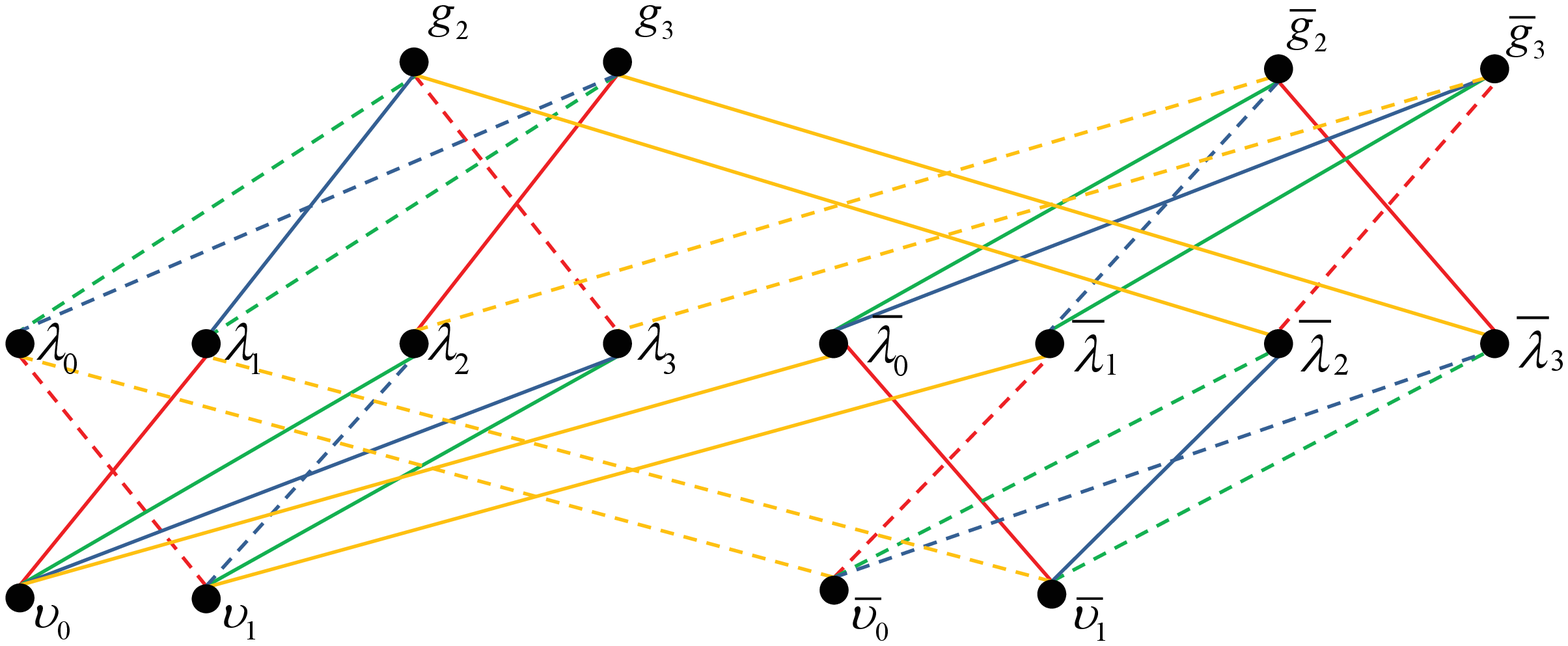}
\end{figure}
\begin{figure}[h]
  \centering
    \includegraphics[width=0.85\textwidth]{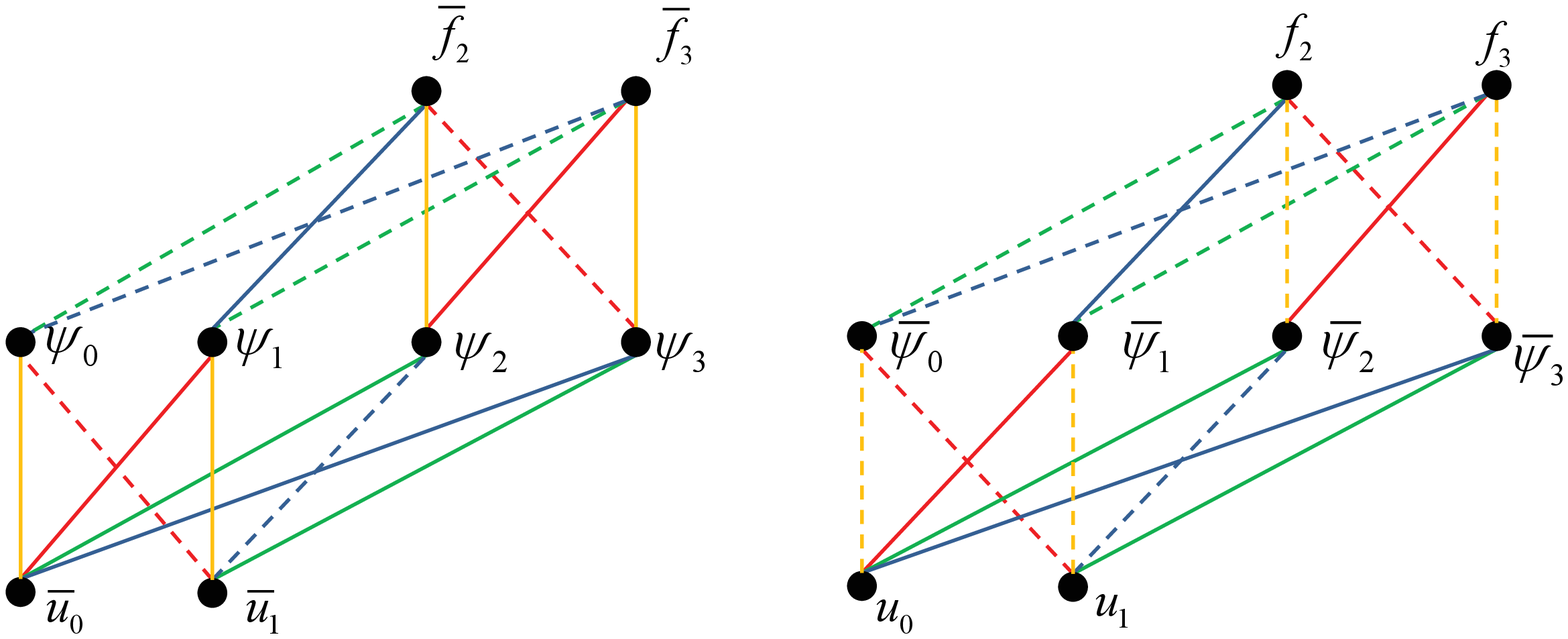}
      \caption{The $(4,8,4)_{D}$ connected graph (above) and the $(4,8,4)_{c,\Delta=0}$ disconnected graph (below), related by the gordian transformation  $u_{0}=\upsilon_{0}-\bar{\upsilon}_{0}$,
      $u_{1}=\upsilon_{1}-\bar{\upsilon}_{1}$, $\bar{u}_{0}=\upsilon_{0}+\bar{\upsilon}_{0}$ and
      $\bar{u}_{1}=\upsilon_{1}+\bar{\upsilon}_{1}$.}
\end{figure}

\subsection{Non-minimal, connected, pure ${\cal N}=4$ supermultiplets revisited}

In order not to overcount the inequivalent supermultiplets, we have to eliminate from (\ref{conngraph})
the supermultiplets which, under a gordian transformation, can be related to a disconnected graph.
It is convenient to apply the notion of {\em pure, connected supermultiplet} only to those pure supermultiplets
which do not admit any presentation in terms of a disconnected graph. Therefore, as we have seen, a supermultiplet with a connected graph
is not necessarily, according to this definition, a connected supermultiplet.\par
The non-minimal, connected, pure ${\cal N}=4$ supermultiplets have length $l=3,4,5$ (the length-$2$ root supermultiplet $(8,8)$ is not connected, see
(\ref{conndiscon})). The connected supermultiplet of length $l=4,5$ are given in (\ref{l45}) . It is indeed easily proved that no gordian transformation
can transform them into a disconnected graph with same length and field content.\par
The complete list of (dually related under mirror symmetry, see the Appendix) non-minimal connected supermultiplets of length $l=3$ is the restriction of (\ref{conngraph}) given by
\begin{eqnarray}\label{connl3}
\begin{array}{|lcl|}\hline
  (1,8,7)_{conn}  & \leftrightarrow & (7,8,1)_{conn} \\ \hline
  (2,8,6)_{A} & \leftrightarrow &(6,8,2)_{A}\\ \hline
  (3,8,5)_{A} & \leftrightarrow & (5,8,3)_{A}\\\hline
  (3,8,5)_{B} & \leftrightarrow & (5,8,3)_{B}\\ \hline
& (4,8,4)_A&\\ \hline
&(4,8,4)_B&\\ \hline
&(4,8,4)_C&\\ \hline
\end{array}
\end{eqnarray}
(the supermultiplets with $(4,8,4)$ field content are self-dual).\par
It is useful to present a further table describing the decompositions of the above
supermultiplets into ${\cal N}=3$ supermultiplets. Indeed,
the supermultiplets in (\ref{connl3}) can be regarded as two minimal ${\cal N}=3$ supermultiplets linked together by
a fourth supersymmetry. Since we have $4$ supersymmetry transformations that can be singled as the ``$4^{th}$" supersymmetry,
there are $4$ ways of decomposing the (\ref{connl3}) supermultiplets into pairs of ${\cal N}=3$ supermultiplets. The following results are
obtained
\begin{eqnarray}
\begin{array}{|ll|}\hline
           & {\cal N}=3\quad decomposition    \\ \hline
      (7,8,1 ): &   {\bf 4\times\{}(3,4,1)+(4,4) \} \\\hline
 (6,8,2 )_A:& {\bf 2\times \{}(3,4,1)+(3,4,1)\}  \\
            &     {\bf 2\times \{}(2,4,2)+(4,4)\} \\\hline
 (5,8,3 )_A: &  {\bf  3\times\{}(3,4,1)+(2,4,2)\}  \\
            &     {\bf 1\times \{}(1,4,3)+(4,4)\} \\\hline
 (5,8,3 )_B: &   {\bf 4\times \{}(3,4,1)+(2,4,2)\} \\\hline
 (4,8,4 )_A: & {\bf 4\times\{}(3,4,1)+(1,4,3)\} \\\hline
 (4,8,4 )_B: & {\bf 3\times\{}(2,4,2)+(2,4,2)\}  \\
            &    {\bf 1\times\{}(4,4,0)+(0,4,4)\} \\\hline
(4,8,4 )_C: &  {\bf 2\times \{}(2,4,2)+(2,4,2)\}  \\
            &     {\bf 2\times\{}(3,4,1)+(1,4,3) \}\\\hline
        \end{array}
\end{eqnarray}
An interesting observation is the following. Two cases, $(4,8,4)_B$ and $(4,8,4)_C$,
produce decompositions into pairs of $(2,4,2)$ ${\cal N}=3$ supermultiplets. For them the  coloring plays an important role.
The $(2,4,2)$ supermultiplets produced from $(4,8,4)_B$ possess the same coloring, while the supermultiplets produced from
$(4,8,4)_C$ possess a different coloring. This property can be stated differently.
Starting from two ${\cal N}=3$ $(2,4,2)$ supermultiplets of the same coloring there exists a unique connected supermultiplet (given by ${\cal N}=4$ $(4,8,4)_B$) that can be obtained by adding a
compatible fourth supersymmetry. Starting from two ${\cal N}=3$ $(2,4,2)$ supermultiplets of different coloring, the unique supermultiplet that can be obtained is given by ${\cal N}=4$ $(4,8,4)_C$.

\section{An entangled ${\cal N}=4$ $(3,8,5)$ supermultiplet}

In \cite{dfghilm3} the possibility of realizing supersymmetry transformations which are not {\em adinkrizable} (in the language of \cite{dfghilm3}), namely that cannot be expressed through a graphical presentation,
was raised.
Explicit examples of supermultiplets which were not described in a way that straightforwardly yielded an Adinkra can be found in \cite{dfghil4} and \cite{dfghil6}, but it was not proven that the component fields could not be rearranged into a form that leads to an Adinkra. This is indeed a hot topic, see e.g. two very recent papers on the existence of non-adinkrizable supermultiplets
\cite{{hk},{ghhs}}. We present here an explicit construction of such type of supermultiplet. We prefer to call it an {\em entangled} supermultiplet. It is constructed by interpolating two adinkrizable supermultiplets ({\em pure supermultiplets},  in our language). We believe that the interpolation provides a natural framework to construct entangled supermultiplets.\par
The specific example of an entangled supermultiplet is given here by interpolating the non-minimal pure ${\cal N}=4$ supermultiplets $(3,8,5)_{b, \Delta=0}$ (a non-chiral disconnected supermultiplet) and
$(3,8,5)_B$ (a connected supermultiplet), presented in Section {\bf 4}.
The four supercharges acting on $(3,8,5)_{b, \Delta=0}$ can be taken as $Q_1,Q_2,Q_3,Q_4$. Four supercharges act on $(3,8,5)_B$, with $Q_5$ replacing $Q_4$. The entangled supermultiplet
$(3,8,5)_\theta$ is constructed in terms of an interpolating angle $\theta$.  The four supercharges acting on it are $Q_1,Q_2,Q_3$ and $Q'=Q_4 \cos\theta+Q_5\sin\theta$.
It is straightforward to show that, for $\theta \neq \frac{n \pi}{2}$, the $(3,8,5)_\theta$ supermultiplet does not admit a graphical presentation (no recombination of the fields with given engineering dimension allows to do that). The pure supermultiplets $(3,8,5)_{b,\Delta=0}$
and $(3,8,5)_B$ are recovered for, respectively, $\theta=0$ and $\theta=\frac{\pi}{2}$.\par
Explicitly, the component fields entering the $(3,8,5)_\theta$ supermultiplet can be
expressed as
\begin{equation}\label{n5thetacomponents}
(\upsilon_{0},\upsilon_{1},\bar{\upsilon}_{0};\lambda_{0},\lambda_{1},\lambda_{2},\lambda_{3},\bar{\lambda}_{0},\bar{\lambda}_{1},\bar{\lambda}_{2},\bar{\lambda}_{3};
g_{2},g_{3},\bar{g}_{1},\bar{g}_{2},\bar{g}_{3}). \nonumber
\end{equation}
The four supersymmetry transformations $Q_{1}, Q_{2}, Q_{3}, Q'$ acting on this set of fields are given by
\begin{eqnarray}
\begin{array}{|c|c|c|c|c|}\hline
                    & Q_{1}       & Q_{2} & Q_{3}                                           & Q'=Q_{4}\cos\theta+Q_{5}\sin\theta \\\hline
  \upsilon_{0}      & \lambda_{1} & \lambda_{2} & \lambda_{3}                               & \lambda_{0}\cos\theta+\bar{\lambda}_{0}\sin\theta \\
  \upsilon_{1}      & -\lambda_{0} & \lambda_{3} & -\lambda_{2}                             & \lambda_{1}\cos\theta+\bar{\lambda}_{1}\sin\theta \\
  \lambda_{0}       & -\dot{\upsilon}_{1} & -g_{2} & -g_{3}                          & \dot{\upsilon}_{0}\cos\theta-\dot{\bar{\upsilon}}_{0}\sin\theta \\
  \lambda_{1}       & \dot{\upsilon}_{0} & -g_{3} & g_{2}                                           & \dot{\upsilon}_{1}\cos\theta-\bar{g}_{1}\sin\theta \\
  \lambda_{2}       & g_{3} & \dot{\upsilon}_{0} & -\dot{\upsilon}_{1}                              & g_{2}\cos\theta-\bar{g}_{2}\sin\theta\\
  \lambda_{3}       & -g_{2} & \dot{\upsilon}_{1} & \dot{\upsilon}_{0}                              & g_{3}\cos\theta-\bar{g}_{3}\sin\theta \\
  g_{2}             & -\dot{\lambda}_{3} & -\dot{\lambda}_{0} & \dot{\lambda}_{1}                   & \dot{\lambda}_{2}\cos\theta+\dot{\bar{\lambda}}_{2}\sin\theta \\
  g_{3}             & \dot{\lambda}_{2} & -\dot{\lambda}_{1} & -\dot{\lambda}_{0}                  & \dot{\lambda}_{3}\cos\theta+\dot{\bar{\lambda}}_{3}\sin\theta \\ \hline
  \bar{\upsilon}_{0} & -\bar{\lambda}_{1} & -\bar{\lambda}_{2} & -\bar{\lambda}_{3}                & \bar{\lambda}_{0}\cos\theta-\lambda_{0}\sin\theta \\
  \bar{\lambda}_{0} & \bar{g}_{1} &\bar{g}_{2} & \bar{g}_{3}                                       & \dot{\bar{\upsilon}}_{0}\cos\theta+\dot{\upsilon}_{0}\sin\theta \\
  \bar{\lambda}_{1} & -\dot{\bar{\upsilon}}_{0} &\bar{g}_{3} & -\bar{g}_{2}                        & \bar{g}_{1}\cos\theta+\dot{\upsilon}_{1}\sin\theta \\
  \bar{\lambda}_{2} & -\bar{g}_{3} & -\dot{\bar{\upsilon}}_{0}& \bar{g}_{1}                        & \bar{g}_{2}\cos\theta+g_{2}\sin\theta  \\
  \bar{\lambda}_{3} & \bar{g}_{2} & -\bar{g}_{1} & -\dot{\bar{\upsilon}}_{0}                       & \bar{g}_{3}\cos\theta+g_{3}\sin\theta \\
  \bar{g}_{1}       & \dot{\bar{\lambda}}_{0} & -\dot{\bar{\lambda}}_{3}& \dot{\bar{\lambda}}_{2}  & \dot{\bar{\lambda}}_{1}\cos\theta-\dot{\lambda}_{1}\sin\theta \\
  \bar{g}_{2}       & \dot{\bar{\lambda}}_{3} & \dot{\bar{\lambda}}_{0} & -\dot{\bar{\lambda}}_{1} & \dot{\bar{\lambda}}_{2}\cos\theta-\dot{\lambda}_{2}\sin\theta\\
  \bar{g}_{3}       & -\dot{\bar{\lambda}}_{2} & \dot{\bar{\lambda}}_{1} & \dot{\bar{\lambda}}_{0} & \dot{\bar{\lambda}}_{3}\cos\theta-\dot{\lambda}_{3}\sin\theta\\\hline
\end{array}
\end{eqnarray}

\section{The $\sigma$-model associated to the ${\cal N}=4$ entangled supermultiplet}

A manifest ${\cal N}=4$ off-shell invariant action can be constructed for the entangled supermultiplet $(3,8,5)_\theta$. The action corresponds to a one-dimensional sigma-model
with a three-dimensional target spanned by the bosonic coordinates $\upsilon_0,\upsilon_1,{\bar {\upsilon }}_0$.
The invariant Lagrangian ${\cal L}$  is constructed, following \cite{{krt},{grt},{gkt}}, through
\begin{equation}\label{n4action}
{\cal L}=Q_{3}Q_{2}Q_{1}Q'F(\upsilon_{0},\upsilon_{1},\bar{\upsilon}_{0}),
\end{equation}
where the four supercharges act as odd Leibniz derivatives on an unconstrained function $F$
(the prepotential) of the three target coordinates. The Lagrangian has the correct dimension for a kinetic term. it is explicitly given by
\begin{equation}\label{n4actionbis}
{\cal L}={\cal L}_{1}\cos\theta+{\cal L}_{2}\sin\theta
\end{equation}
with
\begin{eqnarray}
{\cal L}_{1}&=&\Gamma(\dot{\upsilon}_{0}^{2}+\dot{\upsilon}_{1}^{2}+g_{2}^{2}+g_{3}^{2})-\bar{\Gamma}(\dot{\bar{\upsilon}}_{0}^{2}+\bar{g}_{1}^{2}+\bar{g}_{2}^{2}+\bar{g}_{3}^{2})\nonumber \\\nonumber
     &&+\Gamma(\dot{\lambda}_{0}\lambda_{0}+\dot{\lambda}_{1}\lambda_{1}+\dot{\lambda}_{2}\lambda_{2}+\dot{\lambda}_{3}\lambda_{3})-
     \bar{\Gamma}(\dot{\bar{\lambda}}_{0}\bar{\lambda}_{0}+\dot{\bar{\lambda}}_{1}\bar{\lambda}_{1}+\dot{\bar{\lambda}}_{2}\bar{\lambda}_{2}+\dot{\bar{\lambda}}_{3}\bar{\lambda}_{3})  \\\nonumber &&-\Gamma_{0}(\dot{\upsilon}_{1}(\lambda_{1}\lambda_{0}+\lambda_{3}\lambda_{2})+g_{2}(\lambda_{2}\lambda_{0}+\lambda_{1}\lambda_{3})+g_{3}(\lambda_{3}\lambda_{0}+\lambda_{2}\lambda_{1})
     )\\\nonumber
     &&-\bar{\Gamma}_{\bar{0}}(\bar{g}_{1}(\bar{\lambda}_{0}\bar{\lambda}_{1}+\bar{\lambda}_{2}\bar{\lambda}_{3})+\bar{g}_{2}(\bar{\lambda}_{0}\bar{\lambda}_{2}+\bar{\lambda}_{3}\bar{\lambda}_{1})+
     \bar{g}_{3}(\bar{\lambda}_{0}\bar{\lambda}_{3}+\bar{\lambda}_{1}\bar{\lambda}_{2}))       \\\nonumber
     &&-\Gamma_{1}(\dot{\upsilon}_{0}(\lambda_{0}\lambda_{1}+\lambda_{2}\lambda_{3})+g_{2}(\lambda_{3}\lambda_{0}+\lambda_{2}\lambda_{1})+g_{3}(\lambda_{0}\lambda_{2}+\lambda_{3}\lambda_{1})
     )\\\nonumber
     &&-\bar{\Gamma}_{1}(\dot{\upsilon}_{0}(\bar{\lambda}_{1}\bar{\lambda}_{0}+\bar{\lambda}_{2}\bar{\lambda}_{3})+g_{2}(\bar{\lambda}_{1}\bar{\lambda}_{2}+\bar{\lambda}_{3}\bar{\lambda}_{0})+
     g_{3}(\bar{\lambda}_{0}\bar{\lambda}_{2}+\bar{\lambda}_{1}\bar{\lambda}_{3})     \\\nonumber
     &&+\bar{g}_{1}(\bar{\lambda}_{0}\lambda_{0}+\lambda_{1}\bar{\lambda}_{1}+\bar{\lambda}_{2}\lambda_{2}+\bar{\lambda}_{3}\lambda_{3})+\bar{g}_{2}(\bar{\lambda}_{3}\lambda_{0}+\lambda_{3}\bar{\lambda}_{0}+\lambda_{2}\bar{\lambda}_{1}+\lambda_{1}\bar{\lambda}_{2}) \\\nonumber
     &&+\bar{g}_{3}(\lambda_{0}\bar{\lambda}_{2}+\bar{\lambda}_{0}\lambda_{2}+\lambda_{1}\bar{\lambda}_{3}+\lambda_{3}\bar{\lambda}_{1})-\dot{\bar{\upsilon}}_{0}(\bar{\lambda}_{0}\lambda_{1}+\bar{\lambda}_{1}\lambda_{0}+\bar{\lambda}_{3}\lambda_{2}-\bar{\lambda}_{2}\lambda_{3}))\\\nonumber
     &&-\bar{\Gamma}_{0}(\dot{\upsilon}_{1}(\bar{\lambda}_{0}\bar{\lambda}_{1}+\bar{\lambda}_{3}\bar{\lambda}_{2})+g_{2}(\bar{\lambda}_{0}\bar{\lambda}_{2}+\bar{\lambda}_{1}\bar{\lambda}_{3})+g_{3}(\bar{\lambda}_{0}\bar{\lambda}_{3}+\bar{\lambda}_{2}\bar{\lambda}_{1})\\\nonumber
     &&+\bar{g}_{1}(\lambda_{0}\bar{\lambda}_{1}+\lambda_{1}\bar{\lambda}_{0}+\bar{\lambda}_{3}\lambda_{2}+\lambda_{3}
     \bar{\lambda}_{2})+\bar{g}_{2}(\lambda_{0}\bar{\lambda}_{2}+\lambda_{2}\bar{\lambda}_{0}+\lambda_{1}\bar{\lambda}_{3}+
     \bar{\lambda}_{1}\lambda_{3})\\\nonumber
     &&+\bar{g}_{3}(\lambda_{0}\bar{\lambda}_{3}+\lambda_{3}\bar{\lambda}_{0}+\bar{\lambda}_{2}\lambda_{1}+\lambda_{2}\bar{\lambda}_{1})-\dot{\bar{\upsilon}}_{0}(\bar{\lambda}_{0}\lambda_{0}+\bar{\lambda}_{1}\lambda_{1}+\bar{\lambda}_{2}\lambda_{2}+\bar{\lambda}_{3}\lambda_{3}))\\\nonumber
     &&+\Gamma_{\bar{0}}(\dot{\upsilon}_{1}(\bar{\lambda}_{1}\lambda_{0}+\bar{\lambda}_{0}\bar{\lambda}_{1}+\bar{\lambda}_{3}\lambda_{2}+\lambda_{3}\bar{\lambda}_{2})+g_{2}(\bar{\lambda}_{2}\lambda_{0}+\bar{\lambda}_{0}\lambda_{2}+\bar{\lambda}_{1}\lambda_{3}+\lambda_{1}\bar{\lambda}_{3})\\\nonumber
     &&+g_{3}(\bar{\lambda}_{3}\lambda_{0}+\bar{\lambda}_{0}\lambda_{3}+\bar{\lambda}_{2}\lambda_{1}+\lambda_{2}\bar{\lambda}_{1})-\dot{\upsilon}_{0}(\bar{\lambda}_{0}\lambda_{0}+\bar{\lambda}_{1}\lambda_{1}+\bar{\lambda}_{2}\lambda_{2}+\bar{\lambda}_{3}\lambda_{3})\\\nonumber
     &&+\bar{g}_{1}(\lambda_{0}\lambda_{1}+\lambda_{3}\lambda_{2})+\bar{g}_{2}(\lambda_{0}\lambda_{2}+\lambda_{1}\lambda_{3})+\bar{g}_{3}(\lambda_{0}\lambda_{3}+\lambda_{2}\lambda_{1}))\\\nonumber
     &&+\Gamma_{00}\lambda_{0}\lambda_{1}\lambda_{2}\lambda_{3}-\bar{\Gamma}_{00}(\bar{\lambda}_{1}\lambda_{2}\bar{\lambda}_{3}\lambda_{0}+
     \bar{\lambda}_{0}\lambda_{1}\lambda_{2}\bar{\lambda}_{3}+\bar{\lambda}_{1}\bar{\lambda}_{2}\lambda_{3}\lambda_{0}+\bar{\lambda}_{0}\lambda_{1}\bar{\lambda}_{2}\lambda_{3}) \\\nonumber
     &&+\Gamma_{11}\lambda_{0}\lambda_{1}\lambda_{2}\lambda_{3}-\bar{\Gamma}_{11}(\lambda_{1}\bar{\lambda}_{1}\bar{\lambda}_{3}\lambda_{3}+\lambda_{1}\bar{\lambda}_{1}\bar{\lambda}_{2}\lambda_{2}+
     \lambda_{0}\bar{\lambda}_{0}\bar{\lambda}_{2}\lambda_{2}+\lambda_{0}\bar{\lambda}_{0}\lambda_{3}\bar{\lambda}_{3}) \\\nonumber
     &&+\Gamma_{\bar{0}\bar{0}}(\lambda_{0}\lambda_{1}\bar{\lambda}_{2}\bar{\lambda}_{3}-\bar{\lambda}_{0}\bar{\lambda}_{1}\lambda_{2}\lambda_{3})-
     \bar{\Gamma}_{\bar{0}\bar{0}}(\bar{\lambda}_{0}\bar{\lambda}_{1}\bar{\lambda}_{2}\bar{\lambda}_{3})\\\nonumber
     &&+\Gamma_{0\bar{0}}(\lambda_{1}\bar{\lambda}_{2}\lambda_{3}\lambda_{0}+\bar{\lambda}_{0}\lambda_{1}\lambda_{2}\lambda_{3}+\bar{\lambda}_{1}\lambda_{2}\lambda_{3}\lambda_{0}+\lambda_{1}\lambda_{2}\bar{\lambda}_{3}\lambda_{0})\\\nonumber
     &&-\bar{\Gamma}_{0\bar{0}}(\lambda_{1}\bar{\lambda}_{2}\bar{\lambda}_{3}\bar{\lambda}_{0}+\bar{\lambda}_{1}\bar{\lambda}_{2}\lambda_{3}\bar{\lambda}_{0}+
     \lambda_{0}\bar{\lambda}_{1}\bar{\lambda}_{2}\bar{\lambda}_{3}+\bar{\lambda}_{1}\lambda_{2}\bar{\lambda}_{3}\bar{\lambda}_{0})\\\nonumber
     &&+\Gamma_{1\bar{0}}(\lambda_{1}\lambda_{2}\bar{\lambda}_{2}\lambda_{0}-\lambda_{1}\bar{\lambda}_{1}\lambda_{2}\lambda_{3}+
     \lambda_{0}\bar{\lambda}_{0}\lambda_{2}\lambda_{3}+\lambda_{3}\bar{\lambda}_{3}\lambda_{1}\lambda_{0})\\\nonumber
     &&-\bar{\Gamma}_{1\bar{0}}(\lambda_{2}\bar{\lambda}_{2}\bar{\lambda}_{1}\bar{\lambda}_{0}+\bar{\lambda}_{0}\bar{\lambda}_{2}\bar{\lambda}_{3}\lambda_{0}+
     \lambda_{3}\bar{\lambda}_{3}\bar{\lambda}_{1}\bar{\lambda}_{0}+\bar{\lambda}_{2}\bar{\lambda}_{3}\lambda_{1}\bar{\lambda}_{1})\\\nonumber
     &&+\bar{\Gamma}_{01}(\lambda_{3}\bar{\lambda}_{3}\lambda_{0}\bar{\lambda}_{1}+\lambda_{3}\bar{\lambda}_{3}\lambda_{1}\bar{\lambda}_{0}+
     \lambda_{1}\bar{\lambda}_{1}\bar{\lambda}_{2}\lambda_{3}+\lambda_{2}\bar{\lambda}_{3}\lambda_{1}\bar{\lambda}_{1}\\
     &&+\lambda_{2}\bar{\lambda}_{2}\lambda_{0}\bar{\lambda}_{1}+
     \bar{\lambda}_{0}\lambda_{2}\bar{\lambda}_{3}\lambda_{0}+\lambda_{2}\bar{\lambda}_{2}\lambda_{1}\bar{\lambda}_{0}+\bar{\lambda}_{0}\bar{\lambda}_{2}\lambda_{3}\lambda_{0})
\end{eqnarray}
and
\begin{eqnarray}
{\cal L}_{2}&=&-\Phi(\dot{\upsilon}_{0}^{2}+\dot{\upsilon}_{1}^{2}+\dot{\bar{\upsilon}}_{0}^{2}+g_{2}^{2}+g_{3}^{2}+\bar{g}_{1}^{2}+\bar{g}_{2}^{2}+\bar{g}_{3}^{2})\nonumber\\\nonumber
     &&-\Phi(\dot{\lambda}_{0}\lambda_{0}+\dot{\lambda}_{1}\lambda_{1}+\dot{\lambda}_{2}\lambda_{2}+\dot{\lambda}_{3}\lambda_{3}+
     \dot{\bar{\lambda}}_{0}\bar{\lambda}_{0}+\dot{\bar{\lambda}}_{1}\bar{\lambda}_{1}+\dot{\bar{\lambda}}_{2}\bar{\lambda}_{2}+\dot{\bar{\lambda}}_{3}\bar{\lambda}_{3})\\\nonumber
     &&+ \Phi_{0}(-\dot{\upsilon}_{1}(\bar{\lambda}_{0}\bar{\lambda}_{1}-\lambda_{1}\lambda_{0}-\bar{\lambda}_{2}\bar{\lambda}_{3}+\lambda_{2}\lambda_{3})-g_{2}(\bar{\lambda}_{0}\bar{\lambda}_{2}-
     \lambda_{2}\lambda_{0}-\bar{\lambda}_{3}\bar{\lambda}_{1}+\lambda_{3}\lambda_{1})\\\nonumber
     &&-g_{3}(\bar{\lambda}_{0}\bar{\lambda}_{3}-\lambda_{3}\lambda_{0}-\bar{\lambda}_{1}\bar{\lambda}_{2}+\lambda_{1}\lambda_{2})-\bar{g}_{1}(\lambda_{0}\bar{\lambda}_{1}+
     \lambda_{1}\bar{\lambda}_{0}+\lambda_{3}\bar{\lambda}_{2}-\lambda_{2}\bar{\lambda}_{3})\\\nonumber
     &&-\bar{g}_{2}(\lambda_{0}\bar{\lambda}_{2}+
     \lambda_{2}\bar{\lambda}_{0}+\lambda_{1}\bar{\lambda}_{3}-\lambda_{3}\bar{\lambda}_{1})-\bar{g}_{3}(\lambda_{0}\bar{\lambda}_{3}+
     \lambda_{3}\bar{\lambda}_{0}+\lambda_{2}\bar{\lambda}_{1}-\lambda_{1}\bar{\lambda}_{2})\\\nonumber
     &&+\dot{\bar{\upsilon}}_{0}(-\bar{\lambda}_{0}\lambda_{0}+\bar{\lambda}_{1}\lambda_{1}+\bar{\lambda}_{2}\lambda_{2}+\bar{\lambda}_{3}\lambda_{3}))
     \\\nonumber
     &&+ \Phi_{1}(\bar{g}_{1}(\lambda_{0}\bar{\lambda}_{0}+\lambda_{1}\bar{\lambda}_{1}+\lambda_{2}\bar{\lambda}_{2}+\lambda_{3}\bar{\lambda}_{3})
     -\bar{g}_{2}(\lambda_{1}\bar{\lambda}_{2}+\lambda_{2}\bar{\lambda}_{1}-\lambda_{0}\bar{\lambda}_{3}+\lambda_{3}\bar{\lambda}_{0})
     \\\nonumber&&+\bar{g}_{3}(\lambda_{2}\bar{\lambda}_{0}-\lambda_{0}\bar{\lambda}_{2}-\lambda_{1}\bar{\lambda}_{3}-\lambda_{3}\bar{\lambda}_{1})
     -g_{2}(\lambda_{0}\lambda_{3}+\bar{\lambda}_{3}\bar{\lambda}_{0}+\lambda_{1}\lambda_{2}+\bar{\lambda}_{1}\bar{\lambda}_{2})\\\nonumber
     &&+g_{3}(\lambda_{0}\lambda_{2}+\bar{\lambda}_{3}\bar{\lambda}_{1}+\lambda_{3}\lambda_{1}+\bar{\lambda}_{2}\bar{\lambda}_{0})+
     \dot{\upsilon}_{0}(\lambda_{0}\lambda_{1}+\bar{\lambda}_{0}\bar{\lambda}_{1})+\dot{\bar{\upsilon}}_{0}(\lambda_{3}\bar{\lambda}_{2}-\lambda_{2}\bar{\lambda}_{3}))\\\nonumber
     &&+\Phi_{\bar{0}}(-\bar{g}_{1}(\bar{\lambda}_{0}\bar{\lambda}_{1}-\lambda_{1}\lambda_{0}+\bar{\lambda}_{2}\bar{\lambda}_{3}-\lambda_{2}\lambda_{3})-\bar{g}_{2}(\bar{\lambda}_{0}\bar{\lambda}_{2}-\lambda_{2}\lambda_{0}+\bar{\lambda}_{3}\bar{\lambda}_{1}-\lambda_{3}\lambda_{1})
     \\\nonumber
     &&-\bar{g}_{3}(\bar{\lambda}_{0}\bar{\lambda}_{3}-\lambda_{3}\lambda_{0}+\bar{\lambda}_{1}\bar{\lambda}_{2}-\lambda_{1}\lambda_{2})-g_{2}(\lambda_{1}\bar{\lambda}_{3}-\lambda_{3}\bar{\lambda}_{1}-\lambda_{0}\bar{\lambda}_{2}-\lambda_{2}\bar{\lambda_{0}})  \\\nonumber
     &&-g_{3}(\lambda_{2}\bar{\lambda}_{1}-\lambda_{1}\bar{\lambda}_{2}-\lambda_{0}\bar{\lambda}_{3}-\lambda_{3}\bar{\lambda_{0}})+\dot{\upsilon}_{0}(\bar{\lambda}_{0}\lambda_{0}-\bar{\lambda}_{1}\lambda_{1}-\bar{\lambda}_{2}\lambda_{2}-\bar{\lambda}_{3}\lambda_{3})\\\nonumber
     && -\dot{\upsilon}_{1}(\lambda_{3}\bar{\lambda}_{2}-\lambda_{2}\bar{\lambda}_{3}-\lambda_{0}\bar{\lambda}_{1}-\lambda_{1}\bar{\lambda_{0}}))   \\\nonumber
     &&+\Phi_{00}(\lambda_{1}\lambda_{2}\lambda_{3}\lambda_{0}-\bar{\lambda}_{0}\lambda_{1}\lambda_{2}\bar{\lambda}_{3}+
     \bar{\lambda}_{0}\lambda_{3}\bar{\lambda}_{2}\lambda_{1}-\bar{\lambda}_{0}\bar{\lambda}_{1}\lambda_{2}\lambda_{3})   \\\nonumber
     &&+\Phi_{11}(\lambda_{1}\bar{\lambda}_{1}\lambda_{2}\bar{\lambda}_{2}+\lambda_{0}\bar{\lambda}_{3}\lambda_{2}\lambda_{1}+
     \lambda_{0}\lambda_{3}\bar{\lambda}_{2}\bar{\lambda}_{1}-\bar{\lambda}_{3}\lambda_{3}\lambda_{1}\bar{\lambda}_{1}-
     \lambda_{0}\lambda_{2}\bar{\lambda}_{3}\bar{\lambda}_{1}\\\nonumber
     &&-\bar{\lambda}_{0}\lambda_{0}\bar{\lambda}_{3}\lambda_{3}-
     \bar{\lambda}_{0}\bar{\lambda}_{1}\lambda_{2}\lambda_{3}+\bar{\lambda}_{0}\lambda_{0}\lambda_{2}\bar{\lambda}_{2})  \\\nonumber
     &&+\Phi_{\bar{0}\bar{0}}(\lambda_{0}\lambda_{3}\bar{\lambda}_{2}\bar{\lambda}_{1}+\lambda_{0}\bar{\lambda}_{3}\lambda_{2}\bar{\lambda}_{1}+
     \bar{\lambda}_{0}\bar{\lambda}_{3}\bar{\lambda}_{2}\bar{\lambda}_{1}+\lambda_{0}\bar{\lambda}_{3}\bar{\lambda}_{2}\lambda_{1})  \\\nonumber
     &&+\Phi_{01}(\bar{\lambda}_{2}\lambda_{3}\lambda_{1}\bar{\lambda}_{1}+\bar{\lambda}_{0}\lambda_{0}\lambda_{2}\bar{\lambda}_{3}-
     \bar{\lambda}_{0}\lambda_{2}\bar{\lambda}_{2}\lambda_{1}+\lambda_{1}\bar{\lambda}_{1}\lambda_{2}\bar{\lambda}_{3}+
     \bar{\lambda}_{0}\bar{\lambda}_{3}\lambda_{3}\lambda_{1}\\\nonumber
     &&+\bar{\lambda}_{0}\lambda_{0}\bar{\lambda}_{2}\lambda_{3}+
     \lambda_{0}\lambda_{2}\bar{\lambda}_{2}\bar{\lambda}_{1}-\lambda_{0}\bar{\lambda}_{3}\lambda_{3}\bar{\lambda}_{1})   \\\nonumber
     &&+\Phi_{0\bar{0}}(\lambda_{0}\bar{\lambda}_{1}\lambda_{2}\lambda_{3}+\lambda_{0}\lambda_{1}\bar{\lambda}_{2}\lambda_{3}+
     \bar{\lambda}_{0}\bar{\lambda}_{1}\bar{\lambda}_{2}\lambda_{3}+\bar{\lambda}_{0}\lambda_{1}\bar{\lambda}_{2}\bar{\lambda}_{3}+
     \bar{\lambda}_{0}\bar{\lambda}_{1}\lambda_{2}\bar{\lambda}_{3}\\\nonumber
     &&+\lambda_{0}\lambda_{1}\lambda_{2}\bar{\lambda}_{3}-
     \bar{\lambda}_{0}\lambda_{1}\lambda_{2}\lambda_{3}+\lambda_{0}\bar{\lambda}_{3}\bar{\lambda}_{2}\bar{\lambda}_{1})     \\\nonumber
     &&+\Phi_{1\bar{0}}(\lambda_{1}\bar{\lambda}_{1}\bar{\lambda}_{3}\bar{\lambda}_{2}+\bar{\lambda}_{0}\lambda_{0}\bar{\lambda}_{3}\bar{\lambda}_{2}+
     \lambda_{0}\lambda_{2}\bar{\lambda}_{2}\lambda_{1}+\bar{\lambda}_{0}\lambda_{2}\lambda_{2}\lambda_{1}-\bar{\lambda}_{0}\bar{\lambda}_{3}
     \lambda_{3}\bar{\lambda}_{1}\\\nonumber
     &&-\lambda_{0}\bar{\lambda}_{3}\lambda_{3}\lambda_{1}+\bar{\lambda}_{0}\lambda_{0}\lambda_{2}\lambda_{3}+
     \lambda_{1}\bar{\lambda}_{1}\lambda_{2}\lambda_{3})    \\\nonumber
     &&-(\Gamma+\bar{\Gamma})(\dot{\upsilon}_{0}\dot{\bar{\upsilon}}_{0}+\dot{\upsilon}_{1}\bar{g}_{1}+g_{2}\bar{g}_{2}+g_{3}\bar{g}_{3})\\\nonumber
     &&-(\Gamma+\bar{\Gamma})(\lambda_{0}\dot{\bar{\lambda}}_{0}+\lambda_{1}\dot{\bar{\lambda}}_{1}+\lambda_{2}\dot{\bar{\lambda}}_{2}+\lambda_{3}\dot{\bar{\lambda}}_{3}) \\\nonumber
     &&+(\Gamma+\bar{\Gamma})_{0}(\dot{\upsilon}_{1}\bar{\lambda}_{0}\lambda_{1}+g_{2}\bar{\lambda}_{0}\lambda_{2}+g_{3}\bar{\lambda}_{0}\lambda_{3}
     -\bar{g}_{1}\lambda_{2}\lambda_{3}-\bar{g}_{2}\lambda_{3}\lambda_{1}-\bar{g}_{3}\lambda_{1}\lambda_{2}-
     \dot{\upsilon}_{0}\bar{\lambda}_{0}\lambda_{0})\\ \nonumber
     &&-(\Gamma+\bar{\Gamma})_{\bar{0}}(\dot{\upsilon}_{1}\bar{\lambda}_{2}\bar{\lambda}_{3}+g_{2}\bar{\lambda}_{3}\bar{\lambda}_{1}+g_{3}\bar{\lambda}_{1}\bar{\lambda}_{2}
     +\bar{g}_{1}\bar{\lambda}_{1}\lambda_{0}+\bar{g}_{2}\bar{\lambda}_{2}\lambda_{0}+\bar{g}_{3}\bar{\lambda}_{3}\lambda_{0}\\\nonumber
     &&+\dot{\bar{\upsilon}}_{0}(\bar{\lambda}_{1}\lambda_{1}+\bar{\lambda}_{2}\lambda_{2}+\bar{\lambda}_{3}\lambda_{3}))\\\nonumber
     &&-(\Gamma+\bar{\Gamma})_{1}(g_{2}\lambda_{2}\bar{\lambda}_{1}+g_{3}\lambda_{3}\bar{\lambda}_{1}
     +\bar{g}_{2}\lambda_{0}\lambda_{3}+\bar{g}_{3}\lambda_{2}\lambda_{0}-\dot{\upsilon}_{0}\bar{\lambda}_{1}\lambda_{0}-\dot{\upsilon}_{1}\bar{\lambda}_{1}\lambda_{1}-\dot{\bar{\upsilon}}_{0}\lambda_{2}\lambda_{3})\\\nonumber
        &&+(\Gamma+\bar{\Gamma})_{1\bar{0}}(\lambda_{0}\bar{\lambda}_{1}\bar{\lambda}_{2}\lambda_{2}+\lambda_{0}\bar{\lambda}_{3}
     \lambda_{3}\bar{\lambda}_{1})  \\\nonumber
     &&+(\Gamma+\bar{\Gamma})_{00}\bar{\lambda}_{0}\lambda_{1}\lambda_{2}\lambda_{2}+(\Gamma+\bar{\Gamma})_{11}\lambda_{0}\bar{\lambda}_{1}\lambda_{2}\lambda_{3}
     +(\Gamma+\bar{\Gamma})_{\bar{0}\bar{0}}\lambda_{0}\bar{\lambda}_{1}\bar{\lambda}_{2}\bar{\lambda}_{3}\\
     &&+(\Gamma+\bar{\Gamma})_{01}(\lambda_{0}\bar{\lambda}_{0}\lambda_{2}\lambda_{3}-\lambda_{1}\bar{\lambda}_{1}
\lambda_{2}\lambda_{3}).
\end{eqnarray}
The functions $\Gamma, \bar{\Gamma}$ and $\Phi$ are expressed through the prepotential $F$ as
\begin{equation}
\Gamma={\partial^2}_{00}F+{\partial^2}_{11}F=F_{00}+F_{11},  \qquad  \bar{\Gamma}={\partial^2}_{\bar{0}\bar{0}}F=F_{\bar{0}\bar{0}},\qquad \Phi={\partial^2}_{0\bar{0}}F=F_{0\bar{0}}.
\end{equation}
In the above formulas the partial derivative with respect to $\upsilon_0$ is denoted with the suffix ``$0$" (and similarly for $\upsilon_1$ and $\upsilon_{\bar{0}}$).\par
By setting equal to zero all the fermionic fields in the action we
obtain the bosonic part ${\cal L}_{bos}$ of the Lagrangian
\begin{eqnarray}
{\cal L}_{bos}&=&[\Gamma(\dot{\upsilon}_{0}^{2}+\dot{\upsilon}_{1}^{2}+g_{2}^{2}+g_{3}^{2})-\bar{\Gamma}(\dot{\bar{\upsilon}}_{0}^{2}+\bar{g}_{1}^{2}+\bar{g}_{2}^{2}+\bar{g}_{3}^{2})]\cos\theta\\\nonumber
     &&-[\Phi(\dot{\upsilon}_{0}^{2}+\dot{\upsilon}_{1}^{2}+\dot{\bar{\upsilon}}_{0}^{2}+g_{2}^{2}+g_{3}^{2}+\bar{g}_{1}^{2}+\bar{g}_{2}^{2}+\bar{g}_{3}^{2})\\\nonumber
     &&+ (\Gamma+\bar{\Gamma})(\dot{\upsilon}_{0}\dot{\bar{\upsilon}}_{0}+\dot{\upsilon}_{1}\bar{g}_{1}+g_{2}\bar{g}_{2}+g_{3}\bar{g}_{3}) ]\sin\theta.
\end{eqnarray}
By solving the algebric equations of motion for the auxiliary fields and up to total derivatives we can write
\begin{equation}
{\cal L}_{bos}=g_{ij}\dot{X}^{i}\dot{X}^{j}, \qquad i,j=1,2,3,
\end{equation}
where ${\vec X}=(\upsilon_{0},\upsilon_{1},\bar{\upsilon}_{0})$
and the metric $g_{ij}$ is given by
{
\begin{eqnarray}
g_{ij}&=&\left(
  \begin{array}{ccc}
    \Gamma\cos\theta-\Phi\sin\theta & 0 & -\frac{(\Gamma+\bar{\Gamma})}{2}\sin\theta \\
    0 & (\Gamma-\bar{\Gamma}A^{2})\cos\theta+[(\Gamma+\bar{\Gamma})A-\Phi(1+A^{2})]\sin\theta & 0 \\
    -\frac{(\Gamma+\bar{\Gamma})}{2}\sin\theta & 0 & -\bar{\Gamma}\cos\theta-\Phi\sin\theta\\
  \end{array}
\right),\nonumber\\
&&
\end{eqnarray}
}
with
\begin{equation}
A=\frac{(\Gamma+\bar{\Gamma})\sin\theta}{2(\bar{\Gamma}\cos\theta+\Phi\sin\theta)}.
\end{equation}
The non-vanishing components of the diagonalized metric are
\begin{eqnarray}
g_{11}&=&\frac{1}{2}(\Gamma-\bar{\Gamma})\cos\theta-\Phi\sin\theta+\frac{(\Gamma+\bar{\Gamma})}{2}, \nonumber
\\
g_{22}&=&
(\Gamma-\bar{\Gamma}A^{2})\cos\theta+[(\Gamma+\bar{\Gamma})A-\Phi(1+A^{2})]\sin\theta,\nonumber\\
g_{33}&=&\frac{1}{2}(\Gamma-\bar{\Gamma})\cos\theta-\Phi\sin\theta-\frac{(\Gamma+\bar{\Gamma})}{2}.
\end{eqnarray}
Extra supersymmetry generators can be consistently applied on the $(3,8,5)_\theta$ supermultiplet so that
an ${\cal N}=8$ Extended Supersymmetry can be defined on (\ref{n5thetacomponents}) (in the language of \cite{grt},
the ${\cal N}=5$ $(3,8,5)_\theta$ supermultiplet can be ``oxidized" to an ${\cal N}=8$ supermultiplet).
This property is a consequence of the fact that both $(3,8,5)_{b, \Delta=0}$ and $(3,8,5)_B$ can be ``oxidized" to the same
${\cal N}=8$ $(3,8,5)$ supermultiplet, with supersymmetry generators $Q_I$, $I=1,2,\ldots , 8$. The eight supertransformations acting
on $(3,8,5)_\theta$, besides $Q_1,Q_2,Q_3, Q'$, are $Q'', Q_6,Q_7,Q_8$.  $Q''$ is obtained,
similarly to $Q'$, by rotating the $Q_4,Q_5$ plane
($Q'' =Q_{4}\sin \theta -Q_{5}\cos \theta $).\par
Imposing the ${\cal N}=8$ invariance for the action given by (\ref{n4action}) requires constraining the prepotential $F$. The ${\cal N}=8$
invariance implies the constraint
\begin{eqnarray}\label{n8const}
\Gamma+\bar{\Gamma}&=&0.
\end{eqnarray}
The metric of the ${\cal N}=8$-invariant sigma-model is conformally flat,
\begin{eqnarray}
g_{ij}&=&\delta_{ij}H,
\end{eqnarray}
with the conformal factor $H$ given by
\begin{eqnarray}\label{conformal}
H&=&\Gamma
\cos\theta-\Phi\sin\theta.
\end{eqnarray}
The ${\cal N}=8$-invariant constraint implies that the action given by (\ref{n4actionbis}) is a function of the conformal factor $H$
(the dependence on $\Gamma,\Phi$ enters only through the  conformal factor).

\section{Conclusions}

In this paper we provided the first explicit construction of a supersymmetric one-dimensional sigma-model based on an entangled supermultiplet
(which does not admit a graphical presentation). The possibility of {\em non-adinkrizable}
supermultiplets (here called ``entangled") was raised in \cite{dfghilm3}. Till very recently no explicit example was produced (so that it was even unclear whether this notion could be applied to a non-empty set).  Constructions of non-adinkrizable supermultiplets (in a different context and using different methods from the one proposed here) has been recently discussed in
\cite{{hk,ghhs}}.  Our given example (based on the interpolation between two non-minimal ${\cal N}=4$ supermultiplets of $(3,8,5)$ field content) was suitably chosen to simplify the proof that
there exists no linear combination of the component fields which guarantees a graphical presentation (``Adinkra") of the interpolated supermultiplet in the interval $0<\theta<\frac{\pi}{2}$ of the interpolating angle $\theta$. An important observation is that the interpolating mechanism is a general phenomenon and that entangled supermultiplets tend to proliferate for
large ${\cal N}$ values of the one-dimensional ${\cal N}$-Extended Supersymmetry. It is also important to notice that the entangled supermultiplet has dynamical consequences. An ${\cal N}=4$, one-dimensional, off-shell invariant sigma-model with a three-dimensional target
is based on it. Its action (\ref{n4actionbis}) carries an explicit dependence on $\theta$. This model is supersymmetric  only under the supertransformations specified by the entangled supermultiplet. Therefore, entangled supermultiplets allow to enlarge the class of supersymmetric actions so far considered.\par
In the case of the (\ref{n4actionbis}) action the dependence on $\theta$ can be reabsorbed only
if the constraint (\ref{n8const}), which implies an ${\cal N}=8$ invariance, is imposed.  The
${\cal N}=8$ action turns out to be dependent on the conformal factor (\ref{conformal}).
The ${\cal N}=8$ invariance is made possible by the fact that the two ${\cal N}=4$ pure supermultiplets recovered at $\theta=0$ and $\theta=\frac{\pi}{2}$ can be extended (``oxidized", see \cite{grt}) to the same ${\cal N}=8$ $(3,8,5)$ supermultiplet.
On the other hand, when (\ref{n8const}) is not satisfied, the action (\ref{n4actionbis}) is ${\cal N}=4$ supersymmetric and possesses a genuine $\theta$-dependence.\par
Our investigation about non-minimal pure ${\cal N}=4$ supermultiplets extends the results of \cite{gkt} with the information contained in the node choice group and its possible inequivalent presentations (colorings) under local moves. We have stressed in the Introduction the difference between two types of moves (local and global) acting on graphs and the so-called ``gordian transformations" acting on pure supermultiplets. As a result a given pure supermultiplet can be
associated with inequivalent (under local and global moves) graphs. In certain cases, in particular, a given supermultiplet can be associated to both a disconnected and a connected graph. In order to avoid overcounting, the notion of {\em connected pure supermultiplets} (the supermultiplets which are associated to connected graphs only) has been introduced. The classification of the non-minimal, ${\cal N}=4$, pure, connected supermultiplets has been presented in Section {\bf 3}.\par
The notion of ``coloring", similarly to the notion of ``chiral" supermultiplets \cite{top3}, plays an important role in supersymmetry representations. It is well-known that minimal ${\cal N}=8$ supermultiplets are non-chiral \cite{top3}, being necessarily obtained by linking together (with extra supertransformations) two minimal ${\cal N}=4$ supermultiplets of opposite chirality.
We have shown (see the discussion at the end of Section {\bf 3}) that inequivalent non-minimal ${\cal N}=4$ supermultiplets are obtained by
linking together two ${\cal N}=3$ $(2,4,2)$ supermultiplets based on the fact that their coloring is either the same or different. This property naturally extends to the construction of non-minimal ${\cal N}=5$ supermultiplets by linking together non-minimal ${\cal N}=4$ supermultiplets
(whose respective colorings have been listed here). A forthcoming paper discusses the issues of the representations of the ${\cal N}=5$ supersymmetry. \par
For any given ${\cal N}=4$ supermultiplet, a supersymmetric sigma-model with an ${\cal N}=4$
off-shell invariant action is automatically constructed with the method discussed in \cite{{krt},{grt},{gkt}}
and applied here to the entangled supermultiplet presented in Section {\bf 4}.\par
The representations of the ${\cal N}$-extended one-dimensional supersymmetry and their related off-shell invariant actions are, of course, interesting in their own. An important application, nevertheless, concerns the dimensional reduction of higher-dimensional supersymmetric theories. An $N=2$ supersymmetric theory in a $D=4$ space-time produces
a dimensionally reduced supersymmetric model in one dimension with ${\cal N}=8$
supercharges. A partial spontaneous breaking of supersymmetry to $N=1$ produces, after dimensional reduction, an ${\cal N}=4$-invariant one-dimensional model based on the non-minimal ${\cal N}=4$ supermultiplets. The new feature of the entangled supermultiplets allows to enlarge, as already
stated, the class of admissible supersymmetric models.\par
Let us conclude this paper by pointing out that the present results can be applied to investigate supersymmetry representations in presence of inhomogeneous terms \cite{lt}, non-linear realizations of supersymmetry \cite{{di},{fkt}}, $D$-module representations of superconformal algebras and their associated superconformal mechanics \cite{kt3}. All these extensions (inhomogeneous representations, non-linear realizations, $D$-module representations) are
induced and derived from linear homogeneous supermultiplets, such as those investigated in this work.
\par
\quad\par

{\bf { \Large Appendix: definitions and conventions}}
\par{~}\par

For completeness we report the definitions, applied to the cases used in the text, of the properties
characterizing the homogeneous linear representations of the one-dimensional ${\cal N}$-Extended Superalgebra.
In particular the notions of {\em engineering dimension}, {\em field content}, {\em dressing transformation},
{\em connectivity symbol}, {\em commuting group}, {\em node choice group}, {\em dual supermultiplet} and so on, as well as the association of linear
supersymmetry transformations with graphs, will be reviewed following \cite{pt}-\cite{top3}.
The Reader can consult these papers for broader definitions and more detailed discussions.
A warning is necessary; different names have been given to the same definition, both synchronically (by different groups) and diachronically (they have been modified along the years
in order to avoid possible misunderstandings). Throughout this paper we self-consistently use the definitions below.
 \par
{\em Engineering dimension}:\par
A grading, the engineering dimension $d$, can be assigned to any field entering a linear representation
(the hamiltonian $H$, proportional to the time-derivative operator $\partial\equiv \frac{d}{dt}$,
has a dimension
$1$). Conventionally one can associate bosonic (fermionic) fields with integer (respectively, half-integer) engineering dimension. \par
{\em Field content:}\par
Each finite
linear representation is characterized by its ``field content" \cite{{pt},{krt}}, i.e. the set of integers $(n_1,n_2,\ldots , n_l)$
specifying the number $n_i$ of fields of engineering dimension
$d_i$ ($d_i = d_1 + \frac{i-1}{2}$, with $d_1$ an arbitrary constant) entering the representation.
Physically, the $n_l$ fields of highest dimension are the auxiliary fields which transform as a time-derivative
under any supersymmetry generator. The maximal value $l$ (corresponding to the maximal dimensionality
$d_l$) is defined to be the {\em length} of the representation (a root representation has length $l=2$).
Either $n_1, n_3,\ldots$ correspond to the bosonic fields (therefore $n_2, n_4, \ldots$ specify the fermionic
fields)
or viceversa. \\ In both cases the equality $n_1+n_3+\ldots =n_2+n_4+\ldots = n$ is guaranteed.\par
\par
{\em Dressing transformation:}\par
Higher-length pure supermultiplets are obtained by applying a dressing transformation \cite{pt} to the
length-$2$ root supermultiplet. The root supermultiplet is specified by the ${\cal N}$ supersymmetry
operators ${\widehat Q}_i$ ($i=1,\ldots ,{\cal N})$, expressed in matrix form as
\begin{eqnarray}\label{hatq}
{\widehat Q}_j = \frac{1}{\sqrt{2}}\left(
\begin{array}{cc}
0&\gamma_j\\
-\gamma_j\cdot H&0
\end{array}
\right), &&
{\widehat Q}_{\cal N} =\frac{1}{\sqrt{2}} \left(
\begin{array}{cc}
0&{\bf 1}_n\\
{\bf 1}_n\cdot H&0
\end{array}
\right),
\end{eqnarray}
where  the $\gamma_j$ matrices ($j=1,\ldots , {\cal N}-1$) satisfy the Euclidean Clifford algebra
\bea
\{\gamma_i,\gamma_j\}&=& -2\delta_{ij}{\bf 1}_n.
\eea
The length-$3$ supermultiplets are specified by the ${\cal N}$ operators $Q_i$, given by the dressing transformation
\begin{eqnarray}\label{dressingtransformation}
Q_i &=& D{\widehat Q}_i D^{-1},
\end{eqnarray}
where $D$ is a diagonal dressing matrix such that
\bea\label{dressingmatrix}
{D} &=& \left(
\begin{array}{cc}
{\widetilde{D}}&0\\
0&{\bf 1}_n
\end{array}
\right),
\end{eqnarray}
with ${\widetilde D}$ an $n\times n$ diagonal matrix whose diagonal entries are either $1$ or the derivative
operator $\partial$. \par
{\em Association with graphs:}\par
The association between pure linear supersymmetry transformations and ${\cal N}$-colored oriented graphs \cite{fg} goes as
follows. The fields (bosonic and fermionic) entering a representation
are expressed as vertices. They can be accommodated into an $X-Y$ plane. The $Y$ coordinate
can be chosen to
correspond to the engineering dimension $d$ of the fields. Conventionally, the lowest dimensional fields
can be
associated to vertices lying on the $X$ axis. The higher dimensional fields have positive, integer or
half-integer values of $Y$.
A colored edge links two vertices which are connected by a supersymmetry transformation. Each one of the ${\cal N}$ $Q_i$
supersymmetry generators is associated to a given color. The edges are oriented. The orientation reflects the sign
(positive or negative) of the corresponding supersymmetry transformation connecting the two vertices. Instead
of using
arrows, alternatively, solid or dashed lines can be associated, respectively, to positive or negative signs.
No colored line is drawn for supersymmetry transformations connecting a field with the time-derivative of a
lower
dimensional field. This is in particular true for the auxiliary fields (the fields of highest dimension in the
representation) which are necessarily mapped, under supersymmetry transformations, in the time-derivative
of lower-dimensional fields.
\par
Each {\em pure} irreducible supersymmetry transformation can be presented (the identification is not unique) through
an oriented
${\cal N}$-colored graph with $2n$ vertices. The graph is such that precisely ${\cal N}$ edges, one for each
color, are linked to any given vertex which represents either a $0$-engineering dimension or a $\frac{1}{2}$-engineering
dimension field.  An unoriented ``color-blind" graph can be associated to the initial graph by disregarding
the orientation of the edges
and their colors (all edges are painted in black). \par
{\em Connectivity symbol:}\par
A characterization of length $l=3$ color-blind, unoriented graphs can be expressed through the connectivity
symbol
$\psi_g$ \cite{kt1}, defined as follows
\begin{eqnarray}
\psi_g &=& ({m_1})_{s_1} +({m_2})_{s_2}+\ldots +({m_Z})_{s_Z}.
\end{eqnarray}
The $\psi_g$ symbol encodes the information on the partition of the $n$  $\frac{1}{2}$-engineering dimension fields
(vertices)
into the sets of $m_z$ vertices ($z=1,\ldots, Z$) with $s_z$ edges connecting them to the $n-k$ $1$-engineering
dimension auxiliary fields.
We have
\begin{eqnarray}
m_1+m_2+\ldots +m_Z &=& n,
\end{eqnarray}
while $
s_z\neq s_{z'}$ for $  z\neq z'$.\par
{\em Commuting group:}\par
For a given supermultiplet, its {\em commuting group} \cite{fkt} is the maximal group of linear transformations of the component fields which commute with
all supersymmetry transformations. For a root supermultiplet, its commuting group is read from the Schur's character (real, complex of quaternionic) of the
associated Clifford algebra, see \cite{fkt}. For dressed supermultiplets, the commuting group generators must commute with the dressing operator. The commuting group of the non-minimal supermultiplets has been discussed in the text.\par
{\em Node choice group:}\par
Given a graph associated to an ${\cal N}$-Extended pure supermultiplet, its node choice group
\cite{dfghilm}
is the set of ${\cal N}$-character strings (of $0$'s and $1$'s), closed under the term-by-term ${\bf Z}_2$ addition ($0+0=1+1=0$, $0+1=1+0=1$). An ${\cal N}$-character string of $r$ $1$'s (associated to the supersymmetry generators $Q_{i_1}, \ldots ,Q_{i_r}$) and ${\cal N}-r$ $0$'s (associated to the remaining supersymmetry generators) belongs to the node choice group if and only if for any vertex of the graph (denoted as $V_{in}$), the path $Q_{i_1},\ldots Q_{i_r}$ produces a final vertex $V_{fin}$
with the same engineering dimension as $V_{in}$. Obviously $r$ must necessarily be an even number.\par
A node choice group will be presented either by its set of generators (they will be denoted as ``$<\cdot, \cdot,\ldots>"$), or by its total set of elements (denoted as ``$\{\cdot,\cdot, \ldots\}$"). We have, for instance,
$<1100,0011>\equiv \{0000,1100,0011,1111\}$.\par

{\em Chirality of the ${\cal N}=4$ minimal supermultiplets:}\par
The ${\cal N}=4$ root supermultiplet has a chirality associated with the overall sign ($\eta=\pm 1$)
of the totally antisymmetric tensor $\epsilon_{ijk}$ \cite{top3}. Its supersymmetry transformations are given by
\begin{eqnarray}
Q_{i}(\upsilon_{0},\upsilon_{j};\lambda_{0},\lambda_{j})&=&(\lambda_{i},-\delta_{ij}\lambda_{0}-\eta\epsilon_{ijk}\lambda_{k};-\dot{\upsilon}_{i},\delta_{ij}\dot{\upsilon}_{0}+\eta\epsilon_{ijk}\dot{\upsilon}_{k}),
\\\nonumber
Q_{4}(\upsilon_{0},\upsilon_{j};\lambda_{0},\lambda_{j})&=&(\lambda_{0},\lambda_{j};\dot{\upsilon}_{0},\dot{\upsilon}_{j}),
\end{eqnarray}
for $i,j,k=1,2,3$.\par
The notion of chirality for the root supermultiplet is extended and applied to the chirality of its dressed supermultiplets. \par
For a single supermultiplet the chirality $\eta$ can be flipped via global moves (see the Introduction). For a collection of several ($r=1,2,\ldots, n$) independent ${\cal N}=4$ supermultiplets with chirality
$\eta_r$, the modulus $\Delta=|\sum_r\eta_r |$ is their overall chirality. The modulus is left invariant under local and global moves.
For ${\cal N}=3$ supermultiplets (disregarding $Q_4$) the chirality is not defined because one can flip the sign of $\eta$ with local moves only. \par
{\em Coloring of special supermultiplets:}\par
For the special case of the ${\cal N}=3$ $(2,4,2)$ supermultiplet, its node choice group
admits three different presentations related by global moves:
\\
$NCG_1=<110>\equiv \{110,000\}$,
$NCG_2=<101>\equiv \{101,000\}$,
$NCG_3=<011>\equiv \{011,000\}$.
Let $j\neq 1$ be a third root of unity ($j^3=1$). We can associate to $NCG_i$ the root
$j_i=j^i$, respectively. For a collection of several independent ($r=1,2,\ldots, n$) ${\cal N}=3$ $(2,4,2)$ supermultiplets the modulus $C=|\sum_rj_r|$ is their overall color. Like its chiral counterpart, the modulus $C$ is invariant under both local and global moves.\par
For graphs with ${\cal N}>3$ the notion of coloring is extended to inequivalent (under local moves only)
presentations of its node choice group.

{\em Dual supermultiplet:}\par
A dual supermultiplet is obtained by mirror-reversing, upside-down, the graph associated
to the original supermultiplet.

~
\\ {~}~
\par {\Large{\bf Acknowledgments}}
{}~\par{}~\par
M.G. and K.I. are grateful to CBPF for hospitality. They both acknowledge a PCI-BEV grant. This work was supported by CNPq.

\end{document}